\documentclass[letterpaper, superscriptaddress, FL-netoatfix, twocolumn]{revtex4}
\usepackage{amsmath}
\usepackage{amsfonts}
\usepackage{amssymb}
\usepackage{graphicx}
\usepackage{bbm}
\usepackage{natbib}
\usepackage{hyperref}
\usepackage{color}

\begin{document}

\title{Analytic Continuation by Feature Learning}

\author{Zhe Zhao}
\affiliation{MOE Key Laboratory of Advanced Micro-Structured Materials, School of Physics Science and Engineering, Tongji University, Shanghai 200092, China}

\author{Jingping Xu}
\affiliation{MOE Key Laboratory of Advanced Micro-Structured Materials, School of Physics Science and Engineering, Tongji University, Shanghai 200092, China}

\author{Ce Wang}
\email{cewang@tongji.edu.cn}
\affiliation{MOE Key Laboratory of Advanced Micro-Structured Materials, School of Physics Science and Engineering, Tongji University, Shanghai 200092, China}

\author{Yaping Yang}
\affiliation{MOE Key Laboratory of Advanced Micro-Structured Materials, School of Physics Science and Engineering, Tongji University, Shanghai 200092, China}

\begin{abstract}
Analytic continuation aims to reconstruct real-time spectral functions from imaginary-time Green's functions; however, this process is notoriously ill-posed and challenging to solve. We propose a novel neural network architecture, named the Feature Learning Network (FL-net), to enhance the prediction accuracy of spectral functions, achieving an improvement of at least $20\%$ over traditional methods, such as the Maximum Entropy Method (MEM), and previous neural network approaches. Furthermore, we develop an analytical method to evaluate the robustness of the proposed network. Using this method, we demonstrate that increasing the hidden dimensionality of FL-net, while leading to lower loss, results in decreased robustness. Overall, our model provides valuable insights into effectively addressing the complex challenges associated with analytic continuation.
\end{abstract}

\maketitle

\section{Introduction}\label{sec: sec1}
Imaginary-time Green's functions, which can be directly obtained from quantum Monte Carlo simulations, are fundamental in characterizing many-body systems\cite{Gull2010ContinuoustimeMC, Georges1996DynamicalMT}. However, to understand the real-time dynamics and corresponding spectral functions, analytic continuation is required to extract spectral densities and real-time responses from these imaginary-time data\cite{JARRELL1996133, Schollwoeck2010TheDR}.

Mathematically, analytic continuation involves solving a Fredholm integral equation of the first kind to reconstruct the spectral density $A(\Omega)$ on the real axis from the Green’s function $G(i\omega_n)$ defined on the imaginary axis\cite{JARRELL1996133, PhysRevB.44.6011}. The relationship between these quantities can be expressed by the following integral equation\cite{mahan2000many, Kabanikhin+2011}:
\begin{equation}\label{eq1} 
G(i\omega_n) = \int_{-\infty}^{\infty} \frac{A(\Omega)d\Omega}{i\omega_n - \Omega} , 
\end{equation}
where $G(i\omega_n)$ is the Green’s function at Matsubara frequency $\omega_n$, and $A(\Omega)$ is the spectral density to be determined. The ill-conditioned kernel $K(i\omega_n, \Omega) = \frac{1}{i\omega_n - \Omega}$ makes the inverse problem highly sensitive to small errors in the input data, which leads to significant variations in the resulting spectral density. Due to this sensitivity, analytic continuation is classified as an ill-posed problem, with solutions that are often non-unique and susceptible to noise.
\cite{10.1137/1021044}. The key challenge is to obtain stable and reliable solutions. \cite{Bryan1990MaximumEA}. 

Several approaches have been developed to address this challenge. Nevanlinna\cite{PhysRevLett.126.056402} and Padé approximations\cite{PhysRevB.100.214417, Vidberg1977SolvingTE, PhysRevResearch.2.043263} provide efficient rational function approximations, while methods like singular value decomposition (SVD)\cite{PhysRevB.82.165125} and Tikhonov regularization\cite{Hansen1992AnalysisOD, PhysRevB.107.085129} mitigate noise but increase computational complexity. The maximum entropy method (MEM)\cite{Skilling1989, PhysRevE.94.023303, PhysRevB.108.L201107, PhysRevB.100.075137, PhysRevB.92.060509} is popular for smoothing noisy data but struggles with resolving sharp peaks and coherent excitations\cite{PhysRevLett.124.056401}. 

Recently, neural networks have been explored as a promising method for mapping imaginary-time Green's functions to real-time spectral functions. 
By directly learning a mapping from $G(i\omega_n)$ to $A(\Omega)$ using a given dataset, neural networks can outperform traditional methods such as MEM\cite{PhysRevLett.124.056401, Yao_2022, PhysRevB.105.075112, PhysRevB.100.245123}. However, the trained networks often perform poorly when dealing with multi-peak or sharp-peak spectra and struggle under significant noise\cite{Rakin2019RobustSR}. Moreover, previous studies have lacked comprehensive analyses of the robustness of these networks, which limits their broader adoption. 

\begin{figure}[h]
    \centering
    \includegraphics[width=1.0\linewidth]{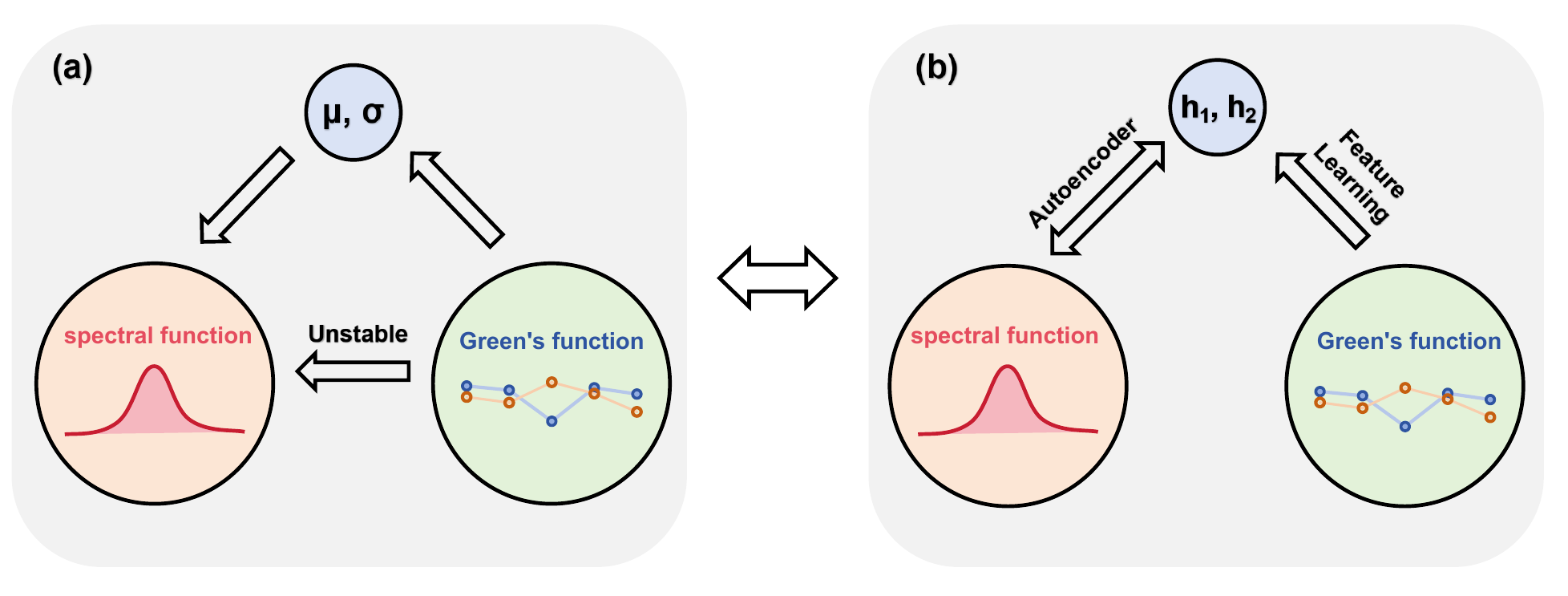}
    \caption{Reconstruction of single-peak spectra: Left—the spectral function is reconstructed by learning its spectral features through feature learning; Right—our neural network reconstructs the spectral function by learning parameterized hidden variables $h$. }
    \label{Com}
\end{figure}

In this paper, we propose a new network architecture that significantly improves the accuracy of spectral function prediction and provides a thorough analysis of its robustness. Our new architecture is founded on a simple observation: the spectral function can always be captured by a few features such as the width, the weight or the position of the peaks. In this sense, although $A$ or $G$ is a high dimensional vector, they can be parameterized by a few features for specific kind of systems. As an instance shown in Fig.~\ref{Com}(a), in single-peak Gaussian spectra, both $A$ and $G$ are uniquely determined by the mean $\mu$ and the variance $\sigma$ of the Gaussian. Hence, though direct mapping from $G$ to $A$ is sensitive to the noise in $G$, another pass from $G$ to $\mu, \sigma$ to $A$ can be more stable. For general data set without any prior knowledge about the features of the spectrum, as shown in Fig.~\ref{Com}(b), similar strategy can be applied by first learning the underlying feature of the data set by an auto-encoder and then training a network that map $G$ to those features.

The rest of this paper is organized as follows. In Section \ref{sec: sec2}, we introduce the design principles of the proposed network architecture and describe the training process, highlighting how the spectral function's key characteristics are leveraged to construct an effective model. In Section \ref{sec: sec3}, we examine the impact of hidden feature dimensionality on model performance through experimental studies and evaluate the effectiveness of different approaches across various datasets. In Section \ref{sec: sec4}, we propose a method utilizing singular value decomposition (SVD) to analyze the robustness of FL-net and compare its robustness across different hidden dimensionalities. Finally, in Section \ref{sec: sec5}, we summarize our findings and propose potential directions for future research.

\section{Network structure}\label{sec: sec2}
\begin{figure}[h]
    \centering
    \includegraphics[width=1.0\linewidth]{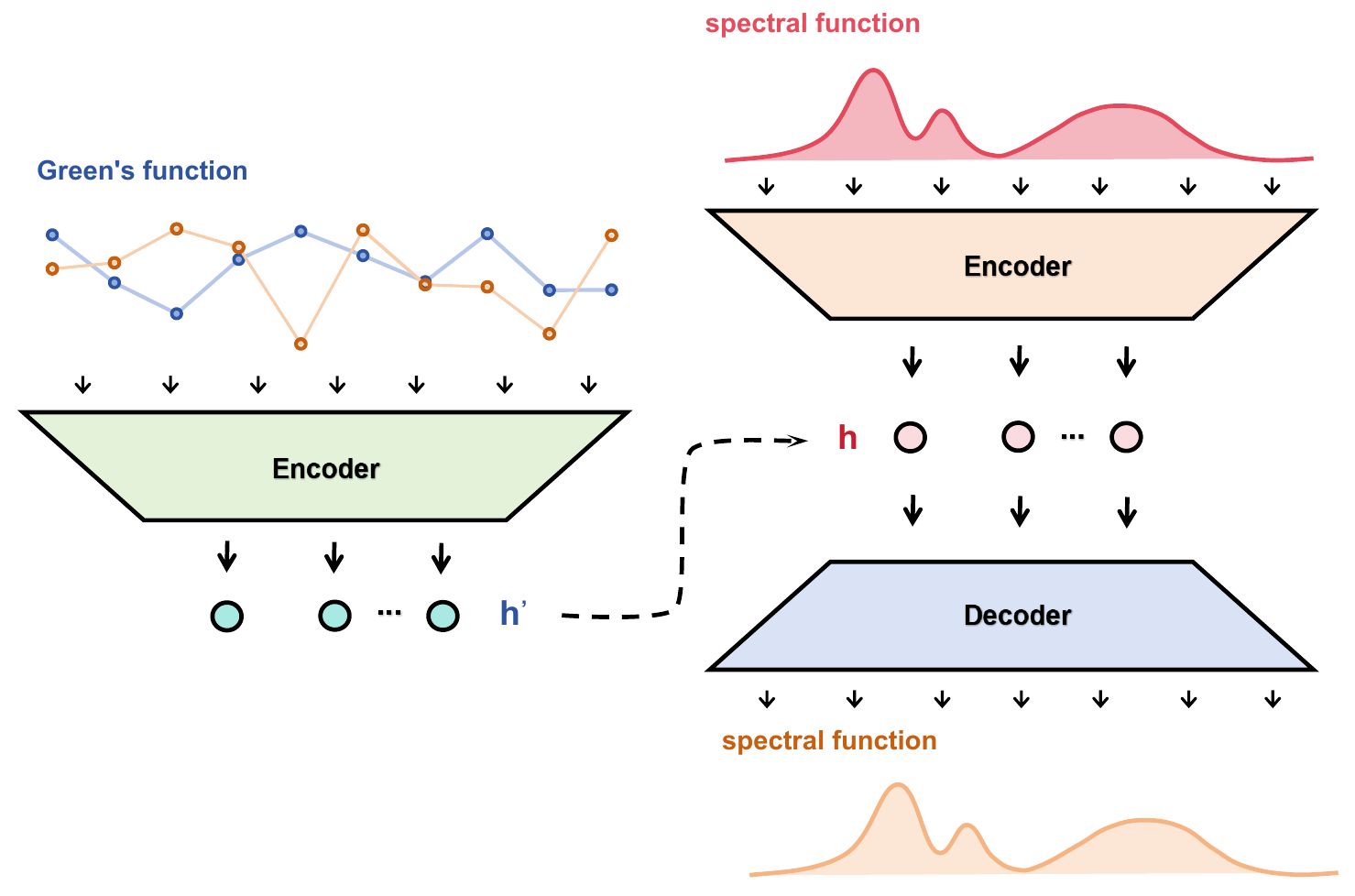}
    \caption{FL-net: The network structure consists of two encoders (one for the imaginary-time Green's function and one for the spectral function) and a decoder. Both the encoder and decoder can have arbitrary architectures. The results consistently demonstrate that the loss from this approach is lower compared to direct learning methods.}
    \label{FL-net}
\end{figure}

We propose the Feature Learning Network (FL-net), as illustrated in Fig.~\ref{FL-net}, based on the described methodology. To prepare $A(\Omega)$ and $G(i\omega_n)$ as input and output for the neural network, we discretize $A(\Omega)$ into a $n_a$ dimensional vector and implement a cutoff $n_g$ on the order of Matsubara frequency $\omega_n$ for $G(i\omega_n)$. 
To be clear, each input data of the spectral function is a vector in the form\cite{Appendix}:
\begin{equation} 
\boldsymbol{A} = \left[ A_1,\, A_2,\, \dots,\, A_{n_a} \right]^\mathrm{T}. 
\end{equation}
The Matsubara Green's function $G(i\omega_n)$ is considered, which is defined at Matsubara frequencies $\omega_n = \frac{(2n + 1)\pi}{\beta}$ for $n = 0, \pm1, \pm2, \dots, \pm(n_g - 1)$, where $\beta = 1/(k_B T)$ is the inverse temperature, with $k_B$ and $T$ denoting Boltzmann's constant and temperature, respectively. At each Matsubara frequency, $G(i\omega_n)$ can be decomposed into its real and imaginary parts: $G_n = \operatorname{Re} G(i\omega_n) + i \operatorname{Im} G(i\omega_n)$. The vector $\boldsymbol{G}$ is formed by arranging these components sequentially: 
\begin{equation} \begin{aligned} 
\boldsymbol{G} = [ & \operatorname{Re} G_{-(n_g - 1)},\, \operatorname{Im} G_{-(n_g - 1)},\, \dots,\, \operatorname{Re} G_{0},\, \operatorname{Im} G_{0}, \\ & \dots,\, \operatorname{Re} G_{n_g - 1},\, \operatorname{Im} G_{n_g - 1} ]^\mathrm{T}. 
\end{aligned} \end{equation}
The training process consisting of three main steps:

First, the auto-encoder consists of two mappings: the encoder $f_A^*: \boldsymbol{A} \to \boldsymbol{h}$ and the decoder $g^*: \boldsymbol{h} \to \boldsymbol{A}$, where both $f_A^*$ and $g^*$ are discrete approximations of the ideal mappings. The latent vector \begin{equation} 
\boldsymbol{h} = [h_1, h_2, \dots, h_{n_h}]^\mathrm{T},
\end{equation}
with adjustable dimension $n_h$, represents the statistical features extracted from $\boldsymbol{A}$. 
 We consider a dataset containing $w$ samples, indexed by $l = 1, 2, \dots, w$. The loss function used to train the auto-encoder is:
\begin{equation} 
\text{MSE}_{A} = \frac{1}{w} \sum_{l=1}^{w} \| \boldsymbol{A}^{(l)} - \hat{\boldsymbol{A}}^{(l)} \|_2^2, 
\end{equation}
where $\boldsymbol{A}^{(l)}$ and $\hat{\boldsymbol{A}}^{(l)}$ denote the true and predicted spectral functions for the $l$-th sample, respectively, and $\| \cdot \|_2$ represents the $L_2$ norm.

Next, we train an encoder performs the approximate mapping $f_G^*: \boldsymbol{G} \to \hat{\boldsymbol{h}}$, where $\hat{\boldsymbol{h}}$ is an approximation of the latent feature $\boldsymbol{h}$. The loss function used for training $f_G^*$ is: 
\begin{equation} 
\text{MSE}_{h} = \frac{1}{w} \sum_{l=1}^{w} \| \boldsymbol{h}^{(l)} - \hat{\boldsymbol{h}}^{(l)} \|_2^2, 
\end{equation}

Finally, we retrain the mapping $g^*: \hat{\boldsymbol{h}} \to \boldsymbol{A}$, completing the transformation $\boldsymbol{G} \to \hat{\boldsymbol{h}} \to \boldsymbol{A}$ \cite{Appendix}. By targeting the latent feature space $\boldsymbol{h}$, FL-net incorporates prior knowledge of the spectral function's peak structures and characteristic distributions.

The network is optimized using the Adam optimizer~\cite{Kingma2014AdamAM}. To balance the influence of each dataset and render the loss dimensionless, we normalize the mean squared error (MSE) by the average variance of the true values. The normalized loss function is defined as:
\begin{equation} 
\mathcal{L} = \frac{\text{MSE}_A}{\sigma_A^2}, 
\end{equation}
where $\sigma_A^2 = \frac{1}{n_a} \sum_{i=1}^{n_a} \mathrm{Var}(A_i)$, and $\mathrm{Var}(A_i)$ represents the variance of the true values of the $i$-th component $A_i$ across the dataset. This normalization mitigates the effect of varying data scales, allowing consistent evaluation across datasets. We use the normalized loss function for all subsequent model evaluations to ensure fair comparisons across datasets.

\addtocounter{figure}{1}

\begin{figure*}[t]
    \centering
    \includegraphics[angle=0,width=8.2cm]{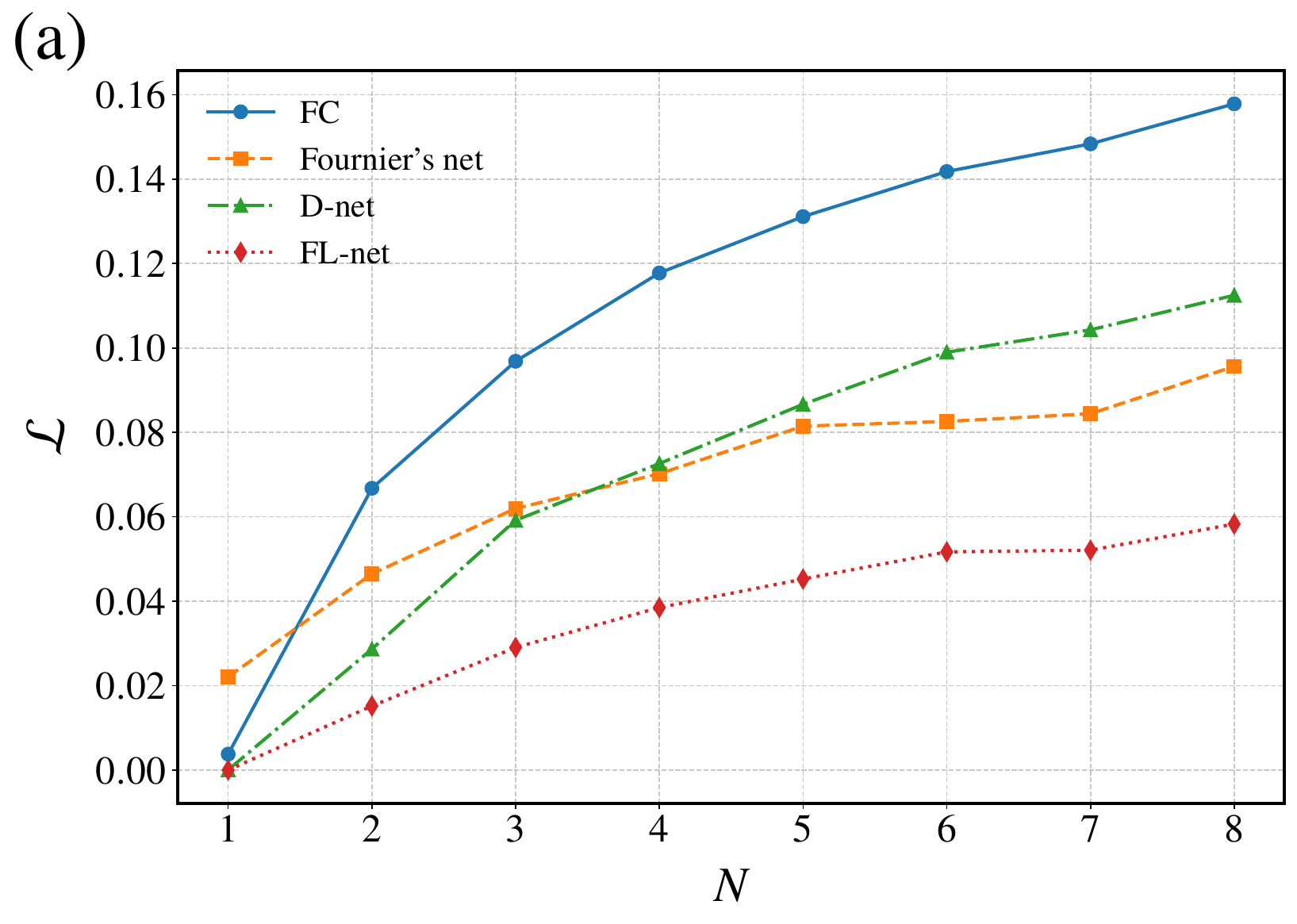}
    \hspace{0.9cm}
    \includegraphics[angle=0,width=7.8cm]{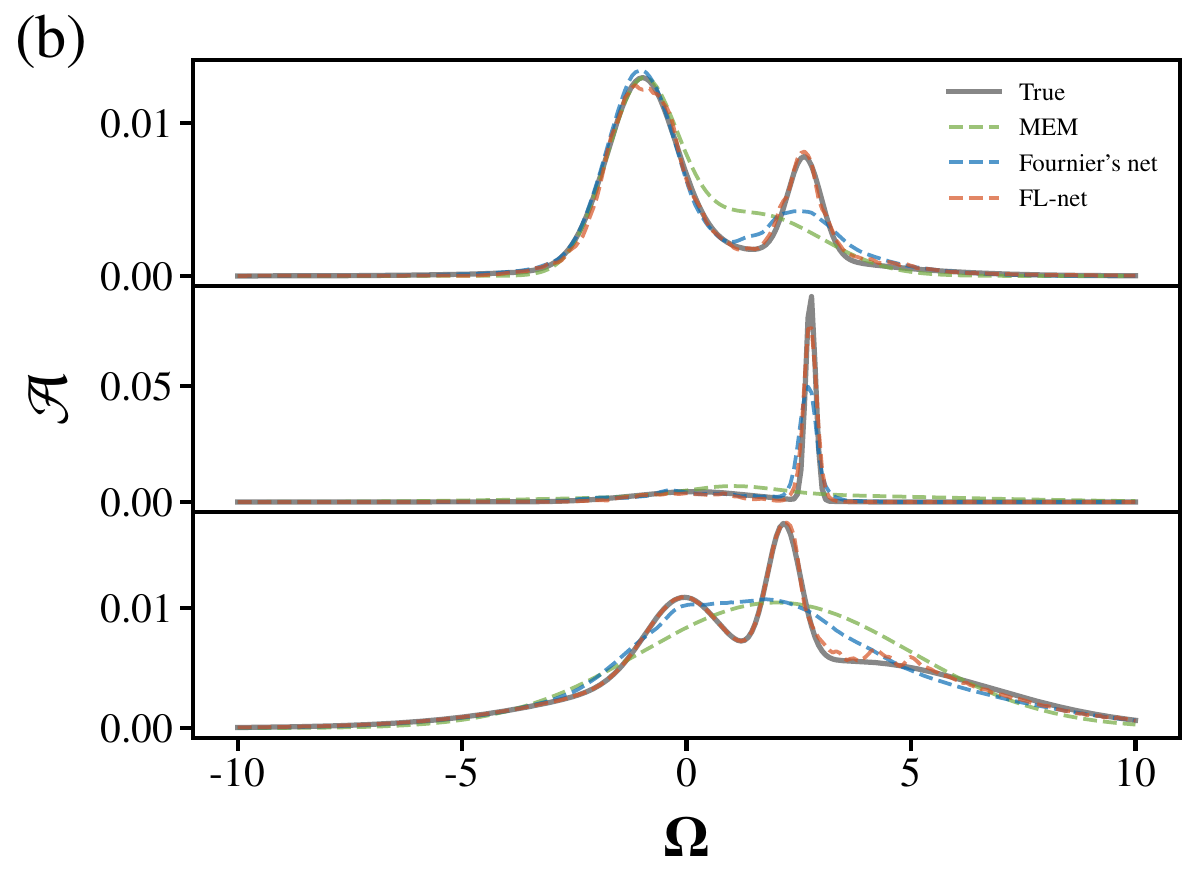}
    \caption{(a) Relative loss comparison across datasets with varying numbers of peaks for different networks: fully connected network (FC net, blue), Fournier network (orange), direct network (D-net, green), and feature learning network (FL-net, red). FL-net achieves the lowest loss, reducing loss by approximately 20$\%$ compared to the Fournier network. (b) Prediction comparison between FL-net (red), Fournier network (blue), and the maximum entropy method (MEM, green) against the ground truth (gray). MEM and the Fournier network struggle with capturing multiple peaks and sharp spectral features, while FL-net consistently matches the true spectra, showing better accuracy in complex cases.}
    \label{result}
\end{figure*}

The code is available for reproduction and further exploration.\footnote{GitHub repository: \url{https://github.com/Order-inz/Analytic-Continuation-by-Feature-Learning}.}

\section{Hidden Feature and Model Performance}\label{sec: sec3}
\addtocounter{figure}{-2}
\begin{figure}[b]
    \centering
    \includegraphics[width=1.0\linewidth]{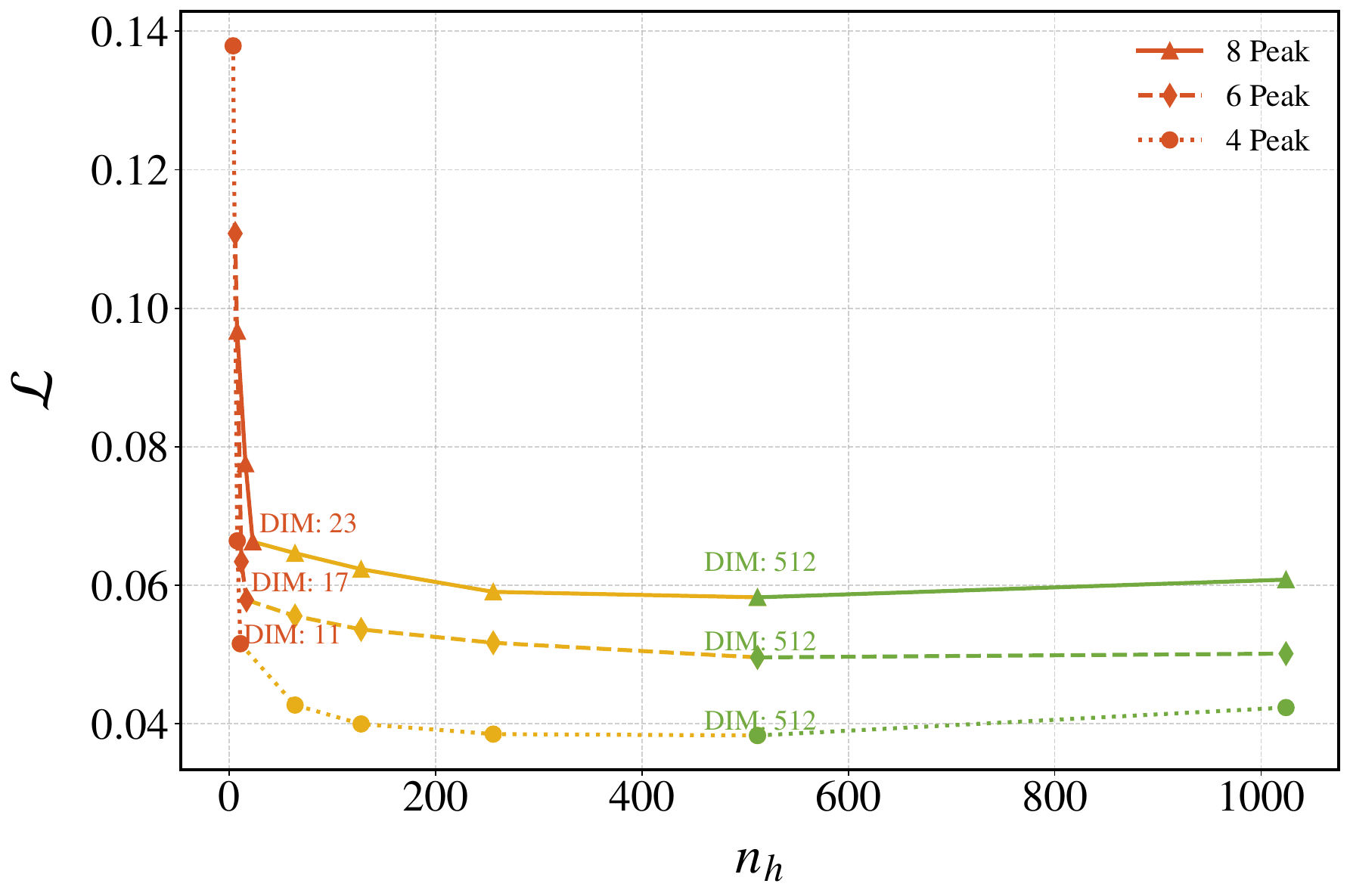}
    \caption{The red curve corresponds to the rapid decrease in loss, demonstrating effective learning with the following hidden dimensions: 8, 16, 23 for 8 peak samples; 6, 12, 17 for 6 peak samples; and 4, 8, 11 for 4 peak samples. The yellow curve represents nodes set to 64, 128, and 256, showing a comparatively slower decline in loss. The green curve, with remaining nodes set to 512 and 1024, shows an increase in loss, suggesting ineffective convergence or overfitting.}
    \label{DIM}
\end{figure}

\addtocounter{figure}{1}
\begin{figure*}[t]
    \centering
    \includegraphics[angle=0,width=4.9cm]{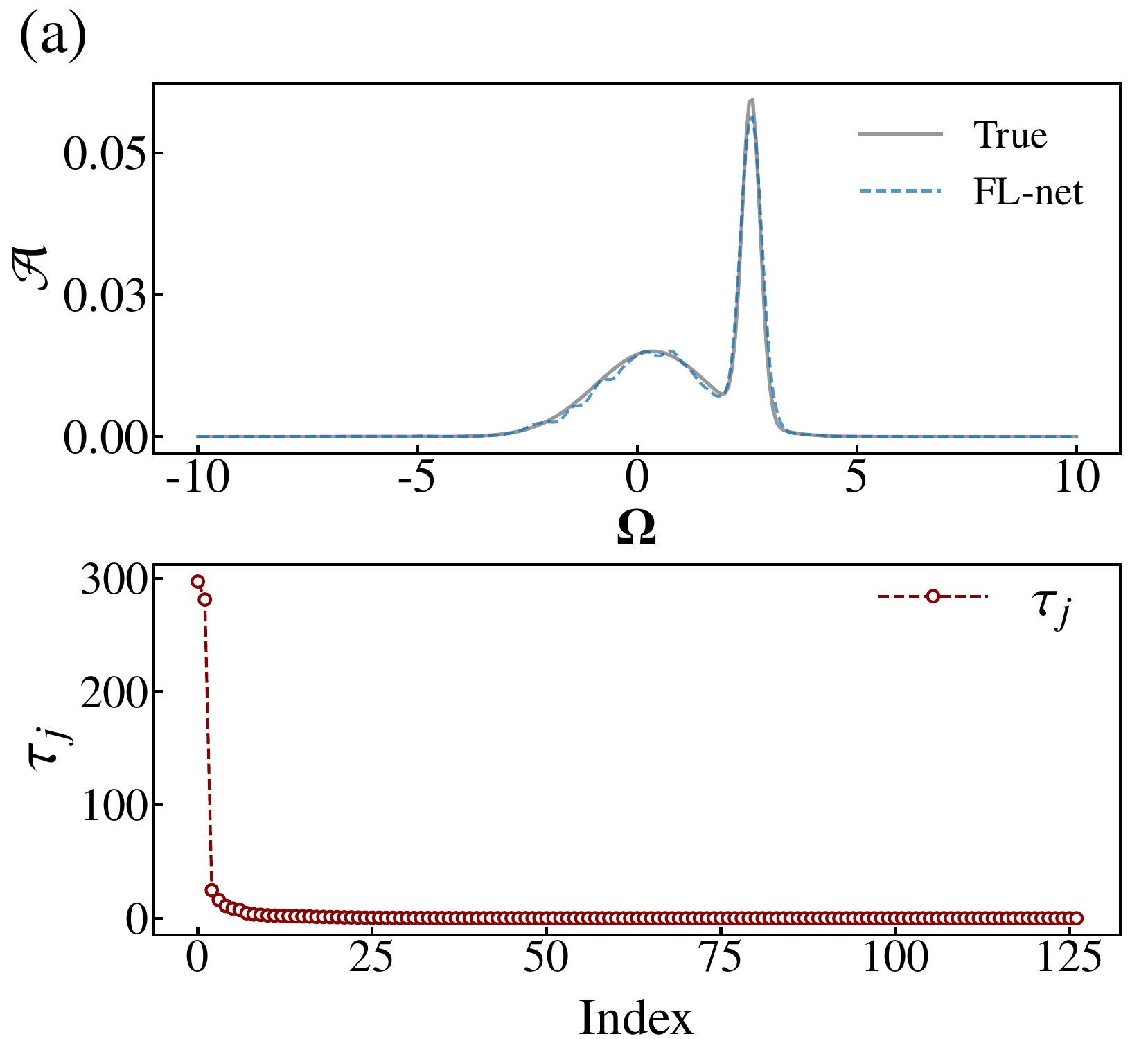}
    \hspace{0.05cm}
    \includegraphics[angle=0,width=6.15cm]{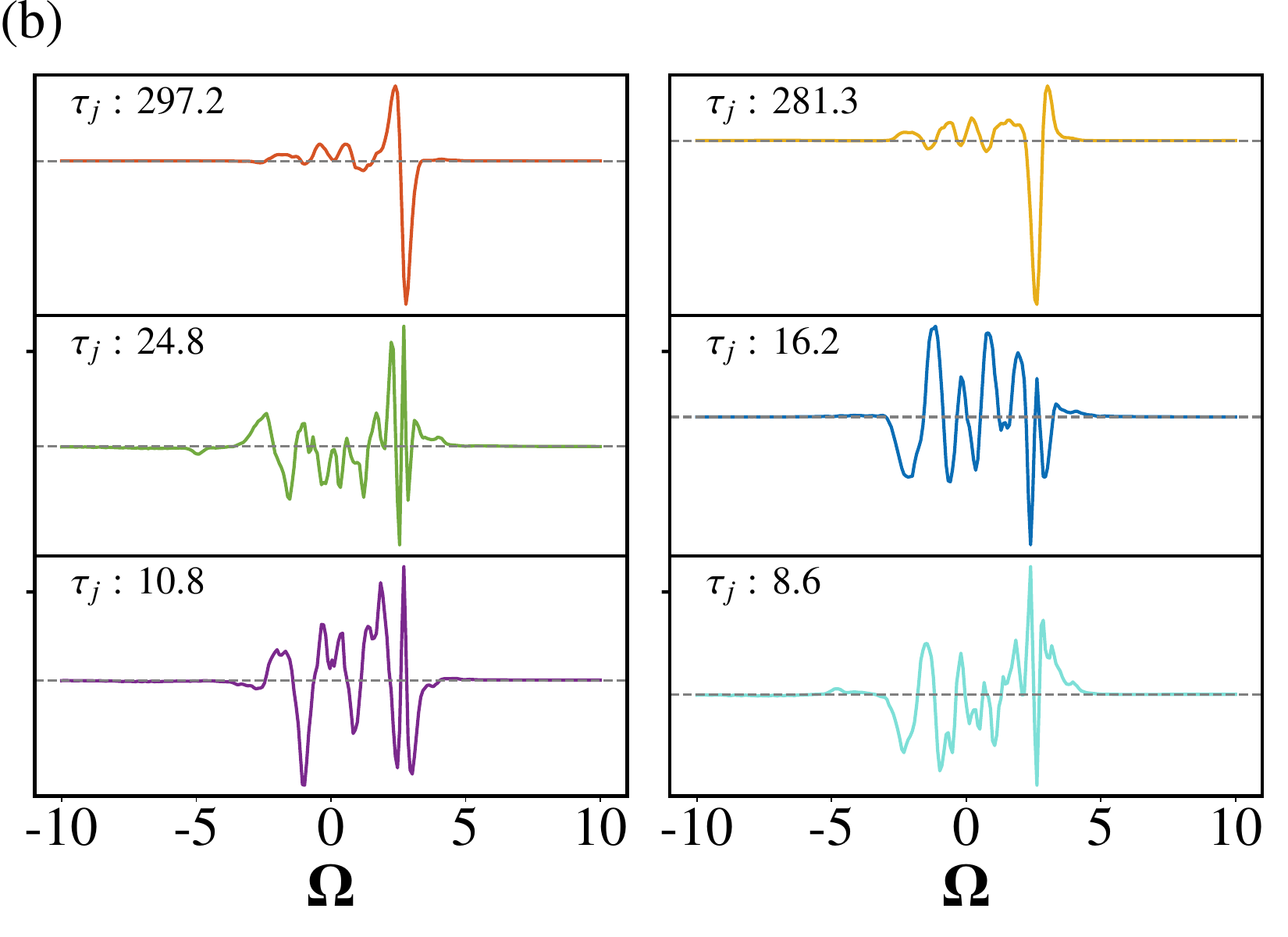}
    \includegraphics[angle=0,width=6.45cm]{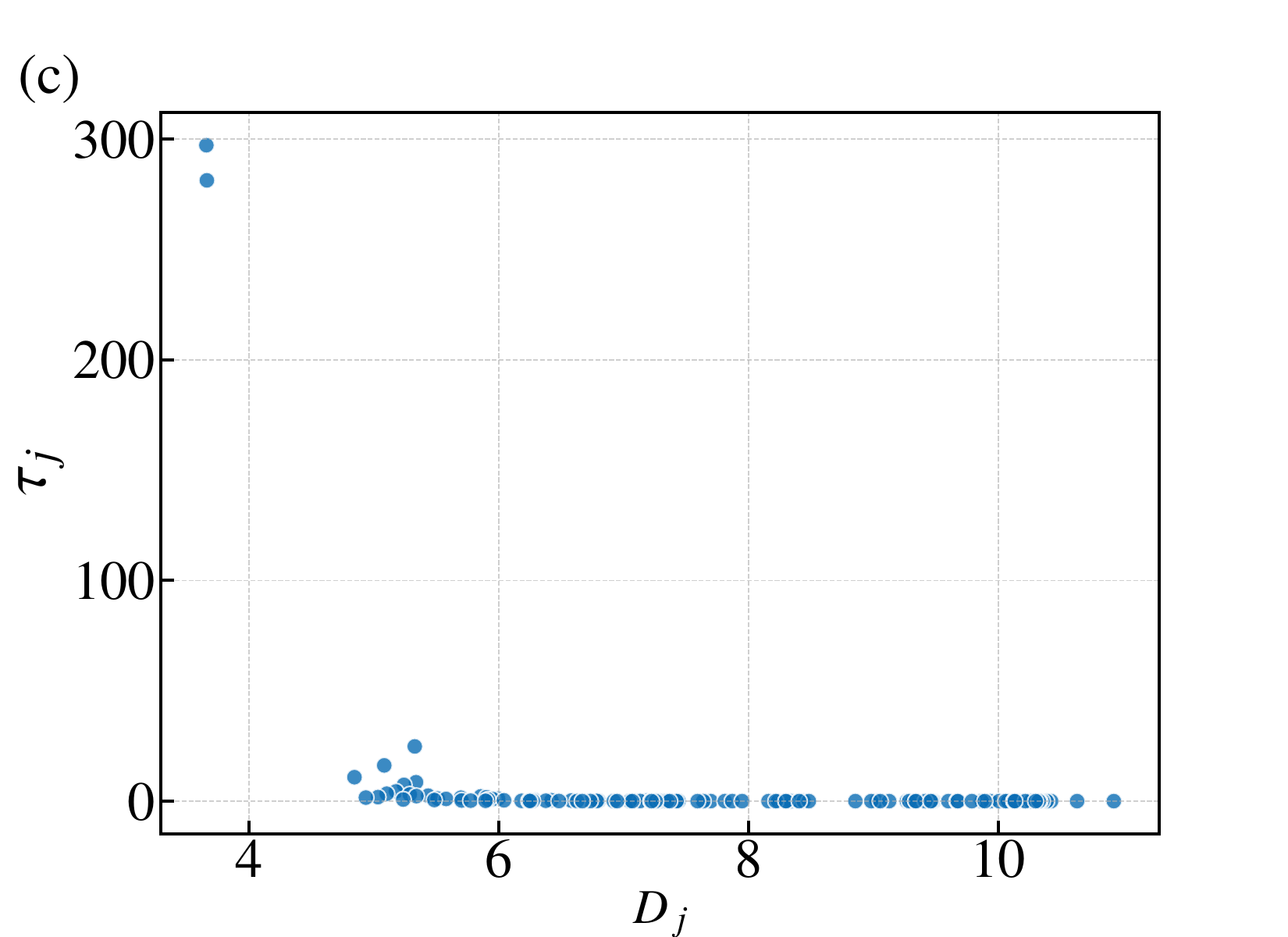}
    \caption{(a) Prediction and singular value: The upper plot shows the true spectral results in gray and the predicted results as a blue dashed line. The lower plot has the horizontal axis representing the matrix column index and the vertical axis representing the corresponding singular values. Most of the column vectors $\boldsymbol{u}_j$ have singular values close to zero, indicating their minimal contribution. (b) The vertical axis shows the mode value and the horizontal axis indicates the mode range. The first six maximum singular values correspond to the left singular vector patterns: Different colors represent different modes, with significant fluctuations in each mode aligning well with features in the predicted results. (c) A scatter plot of the vertical axis $\tau_j$ versus horizontal axis $D_j$ association: Each point represents a mode.}
    \label{SVD}
\end{figure*}

We evaluate the performance of FL-net on a dataset of synthetic spectral functions, each generated as a summation of Gaussian distributions:
\begin{equation}
A(\Omega) = \sum_i \lambda_i \mathcal{N}(\Omega | \mu_i, \sigma_i),
\end{equation}
where each peak is characterized by a mean $\mu_i \in [-5.0, 5.0]$, a standard deviation $\sigma_i \in [0.1, 4.0]$, and $\Omega \in [-10.0, 10.0]$. The parameters $\mu_i$ and $\sigma_i$ are uniformly generated within their respective ranges. The weight coefficients satisfy the normalization condition: $\sum_{i=1}^N \lambda_i = 1$. Using this method, we generate 100,000 spectral functions. The corresponding Green's functions are obtained from the linear mapping described in Eq.~(\ref{eq1}).

We first consider a simple dataset which is consist of single peak Gaussian ($N=1$). By setting the hidden dimension of FL-net to 2, we obtain a considerably small prediction loss on the test set as $\mathcal{L} = 1.02 \times 10^{-4}$. Interestingly, when we use the prior knowledge of Gaussian spectral by setting network mapping process as $\boldsymbol{G} \to \mu, \sigma \to \boldsymbol{A}$ \cite{Appendix}, the prediction loss instead increased to $4.25 \times 10^{-4}$, suggesting that the FL-net has found a more suitable expression for the hidden variable. However, in this single peak Gaussian case, we find the hidden variable $h_1, h_2 $ compressed by FL-net is actually equivalent to $\mu,\sigma$ by numerically computing the Jacobian matrix :

\begin{equation}
    J = 
      \sum_i\begin{pmatrix}
     \frac{\partial h_1}{\partial A_i}\cdot\frac{\partial A_i}{\partial \mu} & \frac{\partial h_1}{\partial A_i}\cdot\frac{\partial A_i}{\partial \sigma} \\ 
     \frac{\partial h_2}{\partial A_i}\cdot\frac{\partial A_i}{\partial \mu} & \frac{\partial h_2}{\partial A_i}\cdot\frac{\partial A_i}{\partial \sigma}
     \end{pmatrix}.
\end{equation}
As an example, the Jacobian matrix evaluated at $\mu = 0$ and $\sigma = 1$ is given by: $J|_{(\mu = 0, \sigma = 1)} = \begin{pmatrix} -0.98 & -0.20 \\ -0.41 & -2.69 \end{pmatrix}$, which is a full rank matrix, indicating that the hidden features $\boldsymbol{h}$ are locally equivalent to $\mu$ and $\sigma$.

For a more general dataset with multi-peak spectra, we begin by investigating how the dimension $n_h$ of the hidden variable $\boldsymbol{h}$ affects the performance of FL-net. As shown in Fig.\ref{DIM}, increasing $n_h$ leads to a rapid decrease in $\mathcal{L}$ of test data until $n_h$ approaches $3N-1$ — the number of independent free parameters in our synthetic data. When $n_h$ exceeds $3N - 1$, the network's loss continues to decrease at a slower rate, yielding only marginal gains. However, further increasing $n_h$ beyond 512 results in an increase in the loss. This performance degradation can be attributed to data sparsity in high-dimensional spaces, which leads to model overfitting and consequently reduces its generalization ability~\cite{Glorot2011DeepSR}. Based on these observations, we will henceforth set $n_h = 256$, striking an optimal balance between performance and computational efficiency.

To verify the effectiveness of the FL-net structure, we introduce a reference network (D-net) with a comparable number of parameters but without utilizing the feature $\boldsymbol{h}$ as the intermediate layer\cite{Appendix}. As illustrated in Fig.~\ref{result}(a), we compare the performance of FL-net and D-net across datasets with varying peak numbers $N$. This comparison clearly shows that incorporating hidden features significantly enhances the accuracy of analytic continuation. Furthermore, we also evaluated FL-net's performance against networks from previous literature\cite{PhysRevLett.124.056401}(Fournier's net) and simple fully connected networks (FC net). The results consistently demonstrated FL-net's superiority in the task of analytic continuation.

We compare the predictions of FL-net, Fournier's net, and the Maximum Entropy Method (MEM) with the ground truth for randomly selected samples, as illustrated in Fig.\ref{result}(b). In the case of double peaks, both MEM and Fournier's net struggle to capture the characteristics of the second peak accurately. For sharp spectra, MEM tends to produce overly smooth results, failing to capture prominent features. As spectral complexity increases, such as in multi-peak scenarios, neither MEM nor Fournier's net adequately captures the complex details. In contrast, FL-net consistently reproduces results closely matching the true spectra, demonstrating superior accuracy in handling complex spectral structures. The strong generalization ability of FL-net on Gaussian datasets suggests its potential for broader applications, including Lorentzian spectra \cite{Appendix}.

\section{Robustness Analysis}\label{sec: sec4}
To examine the robustness of the trained FL-net, we consider the relationship between a small perturbation $\delta \boldsymbol{G}$ in the input imaginary-time Green's function $\boldsymbol{G}$ and the corresponding change $\delta \boldsymbol{A}$ in the output spectral function $\boldsymbol{A}$, which can be expressed as:
\begin{equation}
\delta \boldsymbol{A} = M \delta \boldsymbol{G}, 
\end{equation}
where the matrix $M$ can be naturally obtained from the gradient of the trained neural network. The structure of $M$ can be further analyzed by the singular value decomposition (SVD) :
\begin{equation}
M = U \Sigma V^T,
\end{equation}
where $U$ and $V$ are orthogonal matrices. The matrix $U$ is an orthogonal matrix: 
\begin{equation}
U = \left[ \boldsymbol{u}_1,\, \boldsymbol{u}_2,\, \dots,\, \boldsymbol{u}_{n_a} \right],
\end{equation}
where each $\boldsymbol{u}_j$ is a column vector given by:
\begin{equation}
\boldsymbol{u}_j = \left[ u_{1j},\, u_{2j},\, \dots,\, u_{n_a j} \right]^\mathrm{T}.
\end{equation}
Here, $\boldsymbol{u}_j$ represents the $j$-th eigenmode of $\boldsymbol{A}$ in response to noise in $\boldsymbol{G}$.
In addition, $V$ contains the right singular vectors $\boldsymbol{v}_j$. The diagonal matrix $\Sigma$ contains the singular values $\tau_j$, which quantify the magnitude of $\boldsymbol{A}$'s response to perturbations in $\boldsymbol{G}$. We assume the noise $\delta \boldsymbol{G}$ is totally random, hence the property of the response is determined by $\tau_j$ and $\boldsymbol{u}_j$.

\begin{figure}[t]
    \centering
    \includegraphics[width=1.0\linewidth]{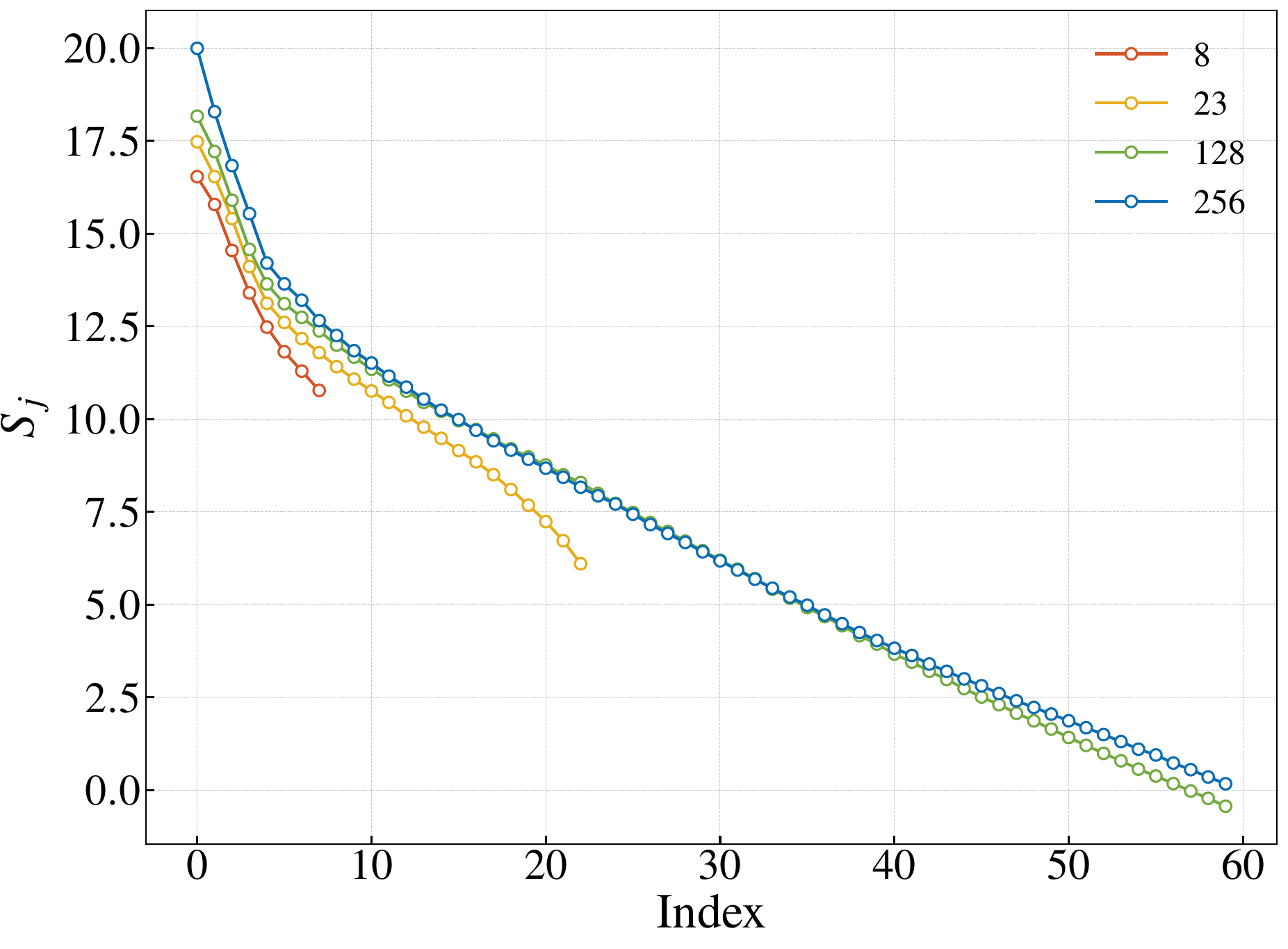}
    \caption{The red, yellow, green, and blue lines represent the robustness performance of networks with hidden dimensions $n_h = 8, 23, 128, 256$, respectively. The vertical axis $S_{j}$ represents the contribution of each mode to the perturbation $\delta A$, while the horizontal axis indicates the index of each column of matrix $U$. As $n_h$ increases, the sensitivity of the primary modes to noise also increases.}
    \label{x}
\end{figure}

The distribution of $\tau_j$ for a fixed $\boldsymbol{A}$ is shown in Fig.\ref{SVD}(a), it appears that there are two modes with significantly larger strength than the remaining ones. However, by further analyzing the pattern of $\boldsymbol{u}_j$ for large $\tau_j$, as illustrated in Fig.\ref{SVD}(b), we find their amplitudes are mainly concentrated at the peak locations of the original spectrum $\boldsymbol{A}$.  In fact, the significant overlap between $\boldsymbol{A}$ and $\boldsymbol{u}_j$ distributions implies that perturbations in the $\boldsymbol{u}_j$ direction only cause small disturbance to $\boldsymbol{A}$. The magnitude of this disturbance is given by the following equation\cite{Appendix}:

\begin{equation} 
D_j = \ln \left( \sum_{i} \frac{u_{ij}^2}{A_i} \right),
\end{equation}
which is the logarithm of the Fisher information along $\boldsymbol{u}_j$ direction. In Fig~\ref{SVD}(c), we plot the singular values $\tau_j$ of each mode against their corresponding $D_j$. It is observed that the two modes with the largest $\tau_j$ are captured by the smallest $D_j$, while the rest modes with small $\tau_j$ have very large $D_j$. Notice that $\tau_j$ represents the overall amplification of the corresponding mode $\boldsymbol{u}_j$. Therefore, the noise sensitivity of each mode should be 

\begin{equation} 
S_{j} = \ln\left( \tau_j^{2} \sum_{i} \frac{u_{ij}^2}{A_i} \right) = 2\ln \tau_j + D_j. 
\end{equation} 


To generally evaluate the robustness of FL-net under different hidden feature dimensions $n_h$, we conducted experiments based on a 1-to-8 peak mixed sample dataset and averaged the $S_j$ values across the dataset. As shown in Fig.~\ref{x}, for networks with lower dimensions, we present the $S_j$ values corresponding to the intermediate hidden layer features; for networks with higher dimensions, we show the top 60 largest $S_j$ values. The analysis results indicate that as $n_h$ increases, $\mathcal{L}$ decreases, whereas the sensitivity of the network to noise increases.



\section{Conclusions}\label{sec: sec5}
In this study, we introduce a novel neural network architecture, the Feature Learning Network (FL-net), designed to learn key statistical properties and essential modes for reconstructing spectral functions. Our results demonstrate that FL-net significantly outperforms previous methods in terms of prediction accuracy. Furthermore, we develop an analytical method to evaluate the robustness of the network. Our analysis reveals that, although increasing the latent feature dimensionality can reduce prediction loss, it may also compromise robustness.

FL-net leverages feature learning to capture the core statistical attributes of spectral functions, providing a new approach for applying machine learning to ill-posed inverse problems. This mechanism has the potential to advance both theoretical understanding and practical applications across a variety of domains. In conclusion, FL-net not only achieves substantial improvements in tackling the analytic continuation problem but also establishes a foundation for applying similar approaches to other complex inverse problems in physics.

\section{Acknowledgement}\label{sec: sec6}
We thank Juan Yao for inspiring discussion. This work is supported by the National Natural Science Foundation of China (NSFC) under Grant Nos. 12204352 (CW).

\appendix
\section{Maximum Entropy Method (MEM)}\label{sec:A}
The classical Maximum Entropy Method is formulated within the framework of Bayesian statistics, as described by Bayes' theorem:
\begin{equation}
P(A|G) = \frac{P(G|A) P(A)}{P(G)},
\end{equation}
where $G$ represents the imaginary-time Green's function obtained through quantum Monte Carlo simulations, and $A$ is the target spectral function. By maximizing $P(A|G)$, we obtain the spectral function that best aligns with the data, while adhering to the prior information encoded in the default model.

In this context, $P(G|A) = e^{-\frac{1}{2}\chi^2}$, where $\chi^2$ quantifies the deviation between the reconstructed and measured Green's functions, defined as:
\begin{equation}
\chi^2 = (G - KA)^{T} C^{-1} (G - KA),
\end{equation}
where $K$ is the transformation matrix mapping the spectral function $A$ to the Green's function $G$, and $C$ is the covariance matrix representing the uncertainties in the data.

The prior probability is given by $P(A) = e^{\alpha S}$, where $S$ is the entropy term of the spectral function, defined as:
\begin{equation}\label{eq14}
S = -\int^{+\infty}_{-\infty} A(\omega) \log\frac{A(\omega)}{D(\omega)} d\omega,
\end{equation}
where $D(\omega)$ is the default model, representing prior knowledge about the spectral function.

Thus, maximizing the objective function $Q = \alpha S - \frac{1}{2}\chi^2$ provides the optimal solution for the spectral function $A$. Once the optimization process converges, the resulting spectral function can be validated against experimental data to assess its reliability.

\section{The details of FL-net, D-net}\label{sec:B}
We discretize the training data for the FL-net by sampling the spectral function $A(\Omega)$ and the Green's function $G(i\omega_n)$ as follows:
For $A(\Omega)$, we sample 256 evenly spaced points over the range $\Omega \in [-10, 10$], forming a 256-dimensional vector: 
\begin{equation} \boldsymbol{A} = [A_1, A_2, \dots, A_{256}]^\mathrm{T}. \end{equation}
For $G(i\omega_n)$, we sample at Matsubara frequencies $\omega_n = \frac{(2n + 1)\pi}{\beta}$ with $n = -32, -31, \dots, 31$, obtaining 64 sample points. We separate the complex values into real and imaginary parts and arrange them sequentially to form a 128-dimensional vector: 
\begin{equation} \boldsymbol{G} = [ \operatorname{Re} G_{-32}, \operatorname{Im} G_{-32}, \dots, \operatorname{Re} G_{31}, \operatorname{Im} G_{31} ]^\mathrm{T}. \end{equation}

\begin{figure}[h]
\centering
\includegraphics[width=1.0\linewidth]{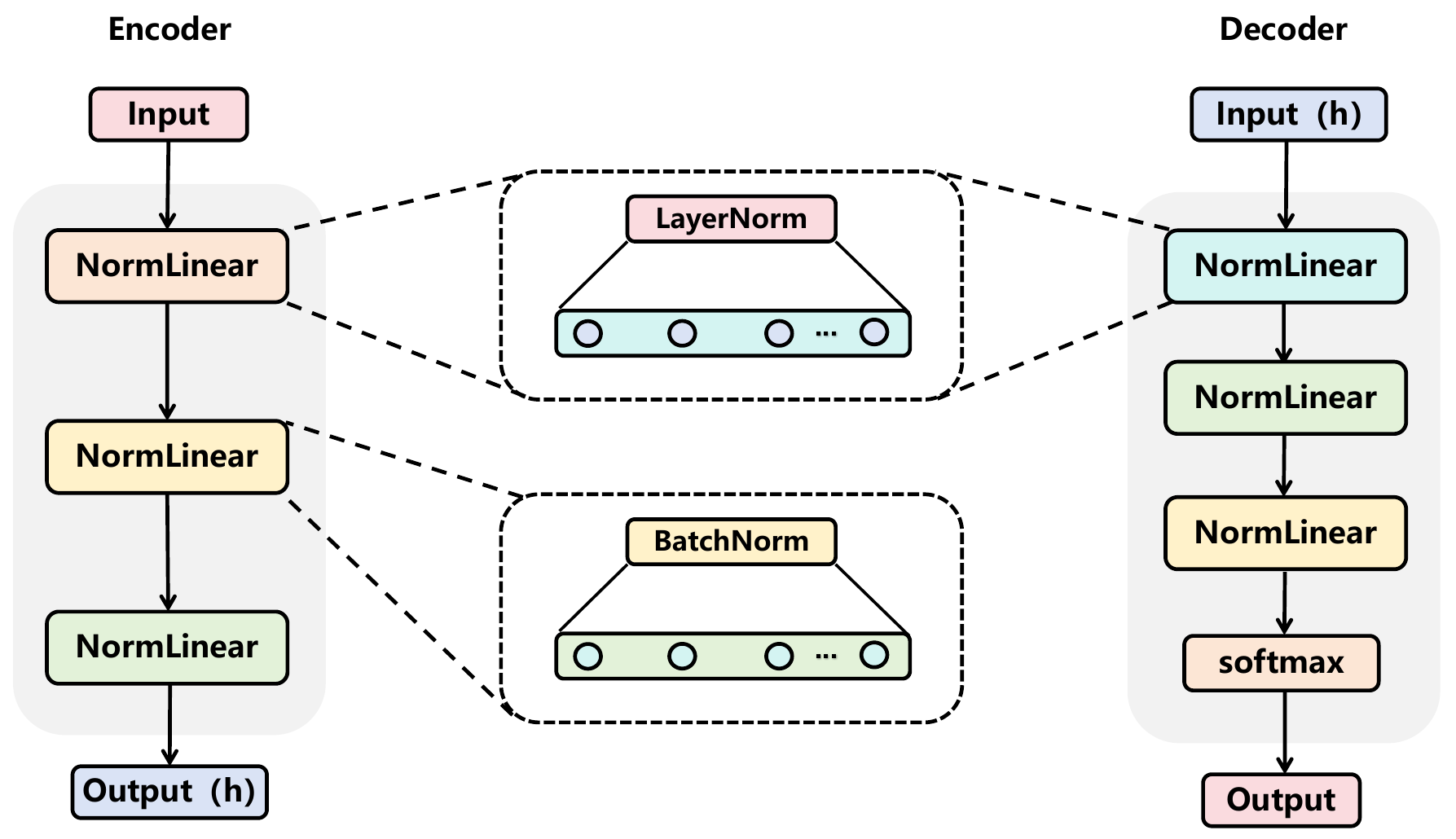}
\caption{Left: Encoder; Right: Decoder. The detailed structure of each module is shown within the dashed boxes.}
\label{En-De}
\end{figure}

The network architecture consists of two encoders and one decoder. The encoders for $\boldsymbol{A}$ and $\boldsymbol{G}$ share the same structure to ensure consistent latent feature distributions (see Fig.~\ref{En-De}). Each encoder primarily uses NormedLinear modules, which include either batch normalization or layer normalization for normalization, and linear layers for feature mapping. The decoder incorporates a softmax layer after the last encoder layer to normalize the predicted spectral function.

For the encoder of the imaginary-time Green's function $\boldsymbol{G}$, the NormedLinear modules utilize batch normalization, which is well-suited for handling Green's functions. In contrast, the encoder for the spectral function $\boldsymbol{A}$ employs layer normalization. Layer normalization is preferred for $\boldsymbol{A}$ because it efficiently manages variations across individual inputs, unlike batch normalization, which may struggle with such variability. By normalizing across layer dimensions and acting on different features within a single sample, layer normalization is more appropriate for analyzing the spectral function\cite{10.5555/3295222.3295349,9382255}.

All layers use the Exponential Linear Unit (ELU) activation function, which maintains non-zero gradients for negative inputs, outperforming ReLU in our experiments\cite{Clevert2015FastAA}.

\begin{figure}[h]
\centering
\includegraphics[width=0.6\linewidth]{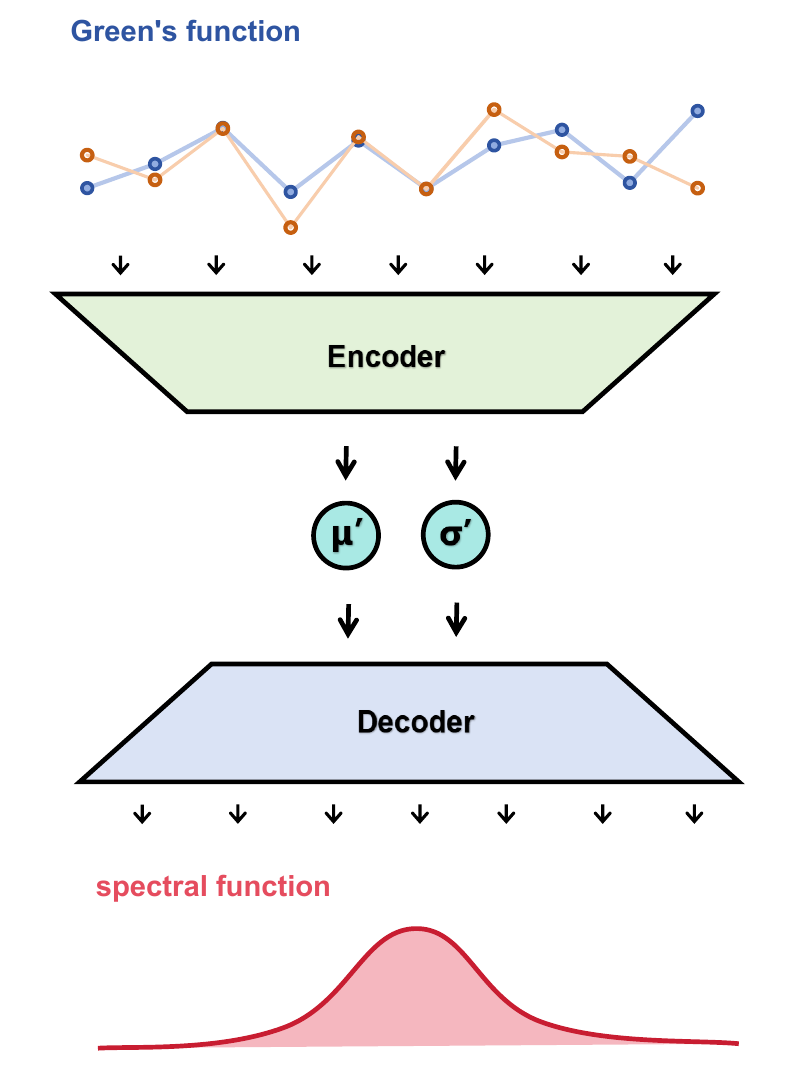}
\caption{Learning process of $\boldsymbol{G} \to \mu',\sigma' \to \boldsymbol{A}$: The encoder and decoder are consistent with those in FL-net.}
\label{toy}
\end{figure}

In the spectral feature learning process, the network trains the encoder by using the Green's function $\boldsymbol{G}$ to directly learn the spectral parameters $\mu$ and $\sigma$, and then uses the encoder to generate new parameters $\mu'$ and $\sigma'$ to reconstruct the spectral function $\boldsymbol{A}$. The goal is to establish a mapping from $\boldsymbol{G} \to \mu', \sigma' \to \boldsymbol{A}$, as illustrated in Fig.~\ref{toy}.

In the D-net, the encoder for $\boldsymbol{G}$ is directly linked to the decoder for $\boldsymbol{A}$, establishing a direct mapping $d^*: \boldsymbol{G} \to \boldsymbol{A}$. In contrast to the FL-net, which uses an intermediate latent space $\boldsymbol{h}$ to capture spectral features, the D-net bypasses this intermediate feature learning. This direct approach enables a clear comparison to assess the effectiveness of the FL-net's feature learning structure.

\section{Results of Lorentzian Visualization}\label{sec:C}
To evaluate the generalizability of the network, we generated Lorentzian-style spectra. The spectral function is defined as:
\begin{equation}
f(\Omega|x_0, \gamma) = \frac{1}{\pi \gamma \left[1+\left(\frac{\Omega-x_0}{\gamma}\right)^2\right]},
\end{equation}
where $x_0 \in [-5, 5]$ represents the peak position, and $\gamma \in [0.1, 3.0]$ is the half-width at half-maximum.

\begin{figure}[h]
\centering
\includegraphics[width=1.0\linewidth]{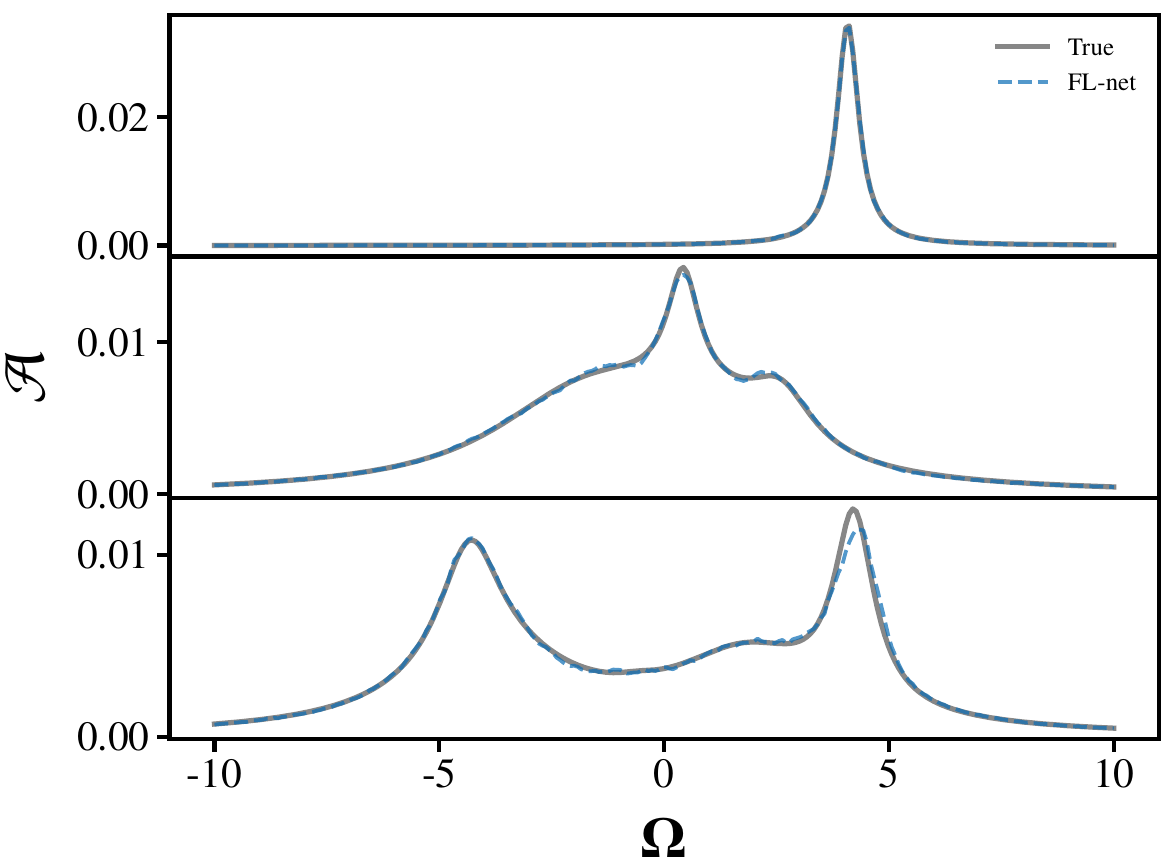}
\caption{The gray line represents the true spectrum, while the blue line shows the prediction from FL-net.}
\label{Lorentzian}
\end{figure}

\begin{table}[h]
    \centering
    \caption{Comparison for Various Methods on Lorentzian}
    \begin{ruledtabular}
    \begin{tabular}{cccccc}
        \textbf{Method} & MEM & FC & Fournier net & Dent & FL-net \\
        \colrule
        \textbf{loss} & 0.1984 & 0.0465 & 0.0244 & 0.0296 & 0.0163 \\
    \end{tabular}
    \end{ruledtabular}
    \label{tab}
\end{table}

Table ref{tab} compares the prediction loss for five different methods on Lorentzian spectra. Our network consistently achieves the lowest loss. As illustrated in Fig.~\ref{Lorentzian}, FL-net demonstrates even greater accuracy in predicting Lorentzian spectra compared to Gaussian spectra.

\section{Details of $D_j$ and $S_j$}\label{sec:D}
Considering that the same perturbation may affect different modes differently, it is essential to analyze the sensitivity of each mode to noise. For this purpose, the output distribution is denoted as $A$, which changes to $A + \delta A$ after a perturbation. To quantify the difference between the distributions before and after the perturbation, we employ the Kullback-Leibler (KL) divergence:

\begin{equation}
D_{KL}(\boldsymbol{A} \parallel \boldsymbol{A} + \delta \boldsymbol{A}) = \sum_i A_i \log\left(\frac{A_i}{A_i + \delta A_i}\right).
\end{equation}

To analyze the response of the KL divergence to a small perturbation $\delta \boldsymbol{A}$, we perform a Taylor expansion. Since $\delta \boldsymbol{A}$ is small and $\sum_i \delta A_i = 0$, the first-order term vanishes, leaving the main contribution from the second-order term:

\begin{equation}
D_{KL}(\boldsymbol{A} \parallel \boldsymbol{A} + \delta \boldsymbol{A}) \approx \frac{1}{2} \sum_{i} \frac{(\delta A_i)^2}{A_i}.
\end{equation}

The perturbation $\delta \boldsymbol{G}$ can be expanded in terms of the right singular vectors $\boldsymbol{v}_j$: $\delta \boldsymbol{G} = \sum_{j} c_j \boldsymbol{v}_j$, where $c_j$ are the projection coefficients. Substituting this into $\delta \boldsymbol{A} = M \delta \boldsymbol{G}$, and noting that $V^\top \boldsymbol{v}_j = \boldsymbol{e}_j$, we obtain:

\begin{equation}
\delta \boldsymbol{A} = U \Sigma V^\top \sum_{j} c_j \boldsymbol{v}_j = U \Sigma \sum_{j} c_j \boldsymbol{e}_j = U \Sigma \boldsymbol{c},
\end{equation}
where $\boldsymbol{c} = [c_1, c_2, \dots, c_{4 (n_g)}]^\mathrm{T}$. Thus, we have:
\begin{equation}
\delta \boldsymbol{A} = \sum_{j} \tau_j c_j \boldsymbol{u}_j.
\end{equation}

To simplify the analysis and focus on the contribution of individual modes, we assume that cross terms can be neglected. This assumption is reasonable if the left singular vectors $\boldsymbol{u}_j$ are approximately orthogonal under the weighting $1/A_i$. This allows us to isolate the contributions of each mode to the KL divergence. Under this assumption, substituting the expression for $\delta A$ into the second-order term of the KL divergence, we obtain:

\begin{equation} \label{eq2}
D_{KL} \approx \frac{1}{2} \sum_{j} \tau_j^2 c_j^2 \sum_{i} \frac{u_{ij}^2}{A_i}.
\end{equation}

In the expression for $D_{\text{KL}}$, the term $\sum_i \frac{u_{ij}^2}{A_i}$ represents the intrinsic sensitivity of each mode to noise, while $\tau_j^2$ indicates the amplification factor of noise in different modes. To measure the inherent sensitivity of each mode, we assume that the coefficients $c_j$ are equal.

This assumption allows us to compare the inherent sensitivities of the modes under uniform perturbation. Therefore, we define the inherent noise sensitivity of the $j$-th mode as:

\begin{equation} 
D_j = \ln \left( \sum_{i} \frac{u_{ij}^2}{A_i} \right),
\end{equation}

To analyze the impact of perturbations on the final output, we consider the amplification effect of the $\tau_j$ term. Thus, we define the total sensitivity $S_j$ of the $j$-th mode as:

\begin{equation} 
S_{j} = \ln\left( \tau_j^{2} \sum_{i} \frac{u_{ij}^2}{A_i} \right) = 2\ln \tau_j + D_j. 
\end{equation}

\bibliographystyle{apsrev4-2}
\bibliography{paper}

\end{document}